\documentclass[11pt,a4paper]{article}

\usepackage{epsfig}
\usepackage{amssymb}
\usepackage{amsmath}
\begin{document}

\title{Braneworld black hole gravitational lens: Strong field limit 
analysis}
\author{Ernesto F. Eiroa\thanks{e-mail: eiroa@iafe.uba.ar} \\
{\small  Instituto de Astronom\'{\i}a y F\'{\i}sica del Espacio, C.C.
67, Suc. 28, 1428, Buenos Aires, Argentina}}
\maketitle
\date{}

\begin{abstract}
In this paper, a braneworld black hole is studied as a gravitational lens, 
using the strong field limit to obtain the positions and magnifications of the 
relativistic images. Standard lensing and retrolensing situations are analyzed 
in a unified setting, and the results are compared with those corresponding to 
the Schwarzschild black hole lens. The possibility of observing the strong 
field images is discussed.
\end{abstract}



\section{Introduction}

Gravitational lensing by ordinary stars and galaxies can be analyzed in the 
weak field approximation, i.e., only keeping the first non null term in the 
expansion of the deflection angle \cite{schne}. If the lens is a black hole, 
this approximation is only valid for photons with large impact parameter, 
so a full strong field treatment is needed instead. In the general case 
where the lens is a compact object with a photon sphere, besides the 
primary and secondary weak field images, two infinite sets of faint 
relativistic images are formed by photons that make complete turns (in both 
directions of rotation) around the black hole before reaching the observer. 
In the last few years, several works studying different strong field lensing 
scenarios appeared in the literature. Virbhadra and Ellis \cite{virbha1} made 
a numerical analysis, using an asymptotically flat metric, of the case where 
the lens is a Schwarzschild black hole situated in the center of the Galaxy, 
and in another paper \cite{virbha2} they investigated numerically the 
lensing by naked singularities. Fritelli, Kling and Newman 
\cite{fritelli} found an exact lens equation without any reference to a 
background metric and compared their results with those of Virbhadra and 
Ellis. In the Schwarzschild geometry, several authors \cite{prev} used a 
logarithmic approximation of the deflection angle as a function of the impact 
parameter, for light rays passing very close to the photon sphere, to treat 
strong field situations. This asymptotic approximation is the starting point 
of an analytical method for strong field lensing, called the strong field 
limit \cite {bozza1}, which gives the lensing observables in a straightforward 
way. Eiroa, Romero and Torres \cite{eiroto} extended this method to 
Reissner-Nordstr\"{o}m geometry, and Bozza \cite{bozza2} showed that it can be 
applied to any static spherically symmetric lens. It was subsequently used by 
Bhadra \cite{bhadra} to study a charged black hole lens of string theory, and 
by Petters \cite{petters} to analyze the relativistic corrections to 
microlensing effects produced by the Galactic black hole. Bozza and Mancini 
\cite{bozman1} applied the strong field limit to study the time delay between 
different relativistic images, showing that different types of black holes are 
characterized by different time delays, and  Bozza \cite{bozza3} extended the 
strong field limit to analyze the case of quasi-equatorial lensing by rotating 
black holes.\\
In standard lensing situations, the lens is placed between the source and the
observer. When the lens has a photon sphere and the observer is placed  
between the source and the lens, or the source is situated between the lens 
and the observer, two infinite sequences of images with deflection angles 
closer to odd multiples of $\pi$ are obtained, a situation called 
retrolensing. Holtz and Wheeler \cite{holtz} recently analyzed the two 
stronger images for a black hole situated in the galactic bulge with the sun 
as source, and suggested retrolensing as a new mechanism for searching 
black holes. De Paolis et al. \cite{depaolis1} considered the retrolensing 
scenario of the bright star S2 orbiting around the massive black hole at the 
galactic center. Eiroa and Torres \cite{eitor} studied the case of a 
spherically symmetric retrolens, using the strong field limit to obtain the 
positions and magnifications of all images.  De Paolis et al. \cite{depaolis2} 
extended, using the strong field limit, the work of Holtz and Wheeler to 
slowly spinning Kerr black hole, restricting their treatment to the black hole 
equatorial plane. Bozza and Mancini \cite{bozman2} analyzed, in the strong 
field limit, standard lensing, retrolensing and intermediate situations under 
a unified formalism, and studied in detail the case of the star S2, suggesting 
the possibility of observing the relativistic images in the year 2018. For 
other works that considered related topics on strong field lensing, see 
Ref. \cite{other}.\\

Braneworld cosmologies, where the ordinary matter is on a three dimensional 
space called the brane, embedded in a larger space called the bulk in which 
only gravity can propagate, became popular in the last few years 
\cite{bwrev}. These models, proposed to solve the hierarchy problem, 
i.e., the difficulty in explaining why the gravity scale is 16 
orders of magnitude greater than the electro-weak scale, have 
motivation in recent developments of string theory, known as 
M-theory. The presence of extra dimensions would affect the characteristics  
of black holes \cite{kanti}. The possibility of the existence of primordial 
black holes in the simplest of braneworld scenarios, the Randall-Sundrum 
\cite{rsII} models (a positive tension brane in a bulk with a negative 
cosmological constant) with one extra dimension, have been studied by 
Clancy, Guedens and Liddle \cite{cgl}. They showed that black holes 
formed in the high energy epoch of this theory have a longer lifetime, 
due to a different evaporation law. Majumdar \cite {majumdar} found 
that the primordial black holes could have a growth of their mass 
through accretion of surrounding radiation during the high energy phase, 
increasing their lifetime. These black holes could have survived up to present 
times and have an induced four dimensional metric on the brane distinct from 
Schwarzschild metric. They also may be created in high energy collisions 
in particle accelerators and in cosmic rays \cite{kanti}. The 
possibility that these primordial black holes 
could act as gravitational lenses was analyzed by Majumdar and Mukherjee 
\cite{majumuk}. They only considered the case of photons with small deflection 
angles in the standard lensing configuration. In another braneworld model 
Frolov, Snajdr and Stojkovic \cite{frolov} calculated, using a weak field 
approximation, the deflection of light propagating in the brane produced by 
a small size black hole in the bulk.\\

In this paper, the strong field limit is applied to study the relativistic 
images produced by the braneworld black hole analyzed by Majumdar and 
Mukherjee \cite{majumuk} only for the weak field situation. Standard lensing 
and retrolensing are considered for the case of high alignment between the 
source, the lens and the observer, which is the case where the images are more 
prominent. In Sec. \ref{s2}, the deflection angle is calculated for the 
braneworld black hole. In Sec. \ref{s3}, the positions and magnifications of 
the relativistic images are obtained, and in Sec. \ref{s4} the results are 
compared with those corresponding to a four dimensional Schwarzschild black 
hole. Finally, in Sec. \ref{s5}, some conclusions are shown. Units such that 
$c=\hbar =1$ are used throughout this work. 

\section{Deflection angle}\label{s2}

Consider a static spherically symmetric black hole that acts as a 
gravitational lens, with metric of the form
\begin{equation}
ds^{2}=-f(x)dt^{2}+g(x)dx^{2}+h(x)d\Omega ^{2},
\label{met}
\end{equation}
where $x=r/r_{H}$ is the radial coordinate in units of the event horizon 
radius. This black hole will have a photon sphere, that corresponds to 
circular unstable photon orbits around it, which radius $x_{ps}$ is 
given by the greater positive solution of the equation:
\begin{equation}
\frac{h^{\prime }(x)}{h(x)}=\frac{f^{\prime }(x)}{f(x)},
\label{ps}
\end{equation}
where the prime means derivative with respect to $x$. Assuming an 
asymptotically flat geometry at infinite, the deflection angle of a photon 
coming from infinite is given as a function of the closest approach distance 
$x_{0}$ by \cite{weinberg}
\begin{equation}
\alpha (x_{0})=I(x_{0})-\pi ,
\label{alp1}
\end{equation}
with
\begin{equation}
I(x_{0})=\int_{x_{0}}^{\infty }2\left[ \frac {g(x)}{h(x)}\right] ^{1/2}
\left[ \frac {h(x)f(x_{0})}{h(x_{0})f(x)}-1\right] ^{-1/2}dx.
\label{i1}
\end{equation}
There are two cases where the integral can be approximated by simple 
expressions. For photons with $x_{0}\gg x_{ps}>1$, which have small 
deflection angles, the integral is usually replaced by a Taylor expansion in 
terms of $1/x_{0}$, keeping only the first non null term. This approximation 
is called the weak field limit, and it is used in lensing by stars and 
galaxies \cite{schne}, and also for the primary and secondary images, in the 
standard lensing configuration, with high alignment, for black hole lenses 
\cite{bozza1,eiroto}. The integral diverges when $x_{0}= x_{ps}$, and for 
$0<x_{0}-x_{ps}\ll 1$, its large value can be asymptotically approximated by 
a logarithmic function \cite{bozza2}. This case, which corresponds to large 
deflection angles, is called the strong field limit \cite{bozza1}. The 
treatment of intermediate situations is more difficult, because one cannot 
rely on simple approximations.\\   

Here it is studied as a gravitational lens, in the strong field limit, the 
braneworld black hole considered as a primordial black hole in Refs. 
\cite{cgl,majumdar}, and as a gravitational lens, in the weak field limit, 
in Ref. \cite{majumuk}. This black hole is nonrotating and it has no charge, 
and its geometry is that of a Schwarzschild solution in five dimensions. The 
Randall-Sundrum II \cite{rsII} braneworld model is adopted, consisting of a 
single positive tension brane with three spatial dimensions, embedded in a 
one dimensional bulk with negative cosmological constant. For the 
event horizon radius $r_{H}$ much smaller than the AdS radius $l$, this black 
hole is a good approximation, in the neighborhood of the event horizon, of a 
black hole formed from collapsed matter confined to the brane. The induced 
four dimensional metric on the brane is \cite{cgl}
\begin{equation}
ds^{2}=-\left( 1-\frac{r_{H}^{2}}{r^{2}}\right)dt^{2}+
\left( 1-\frac{r_{H}^{2}}{r^{2}}\right)^{-1}dr^{2}+r^{2}d\Omega ^{2},
\label{bbh0a}
\end{equation}
with the black hole horizon radius $r_{H}$ given by
\begin{equation}
r_{H}=\sqrt{\frac{8}{3\pi }}\left( \frac{l}{l_{4}}\right)^{1/2} 
\left( \frac{M}{M_{4}}\right)^{1/2}l_{4},
\label{bbh0b}
\end{equation}
where $M$ is the black hole mass, and $l_{4}$ and $M_{4}$ are, 
respectively, the four dimensional Planck length and mass.
Using the radial coordinate $x$ defined above, the metric functions are
$f(x)=g(x)^{-1}=1-x^{-2}$ and $h(x)=x^{2}$. Then
\begin{equation}
I(x_{0})=2x_{0}^{2}\int_{x_{0}}^{\infty }
\left[ x^{4}(x_{0}^{2}-1)-x_{0}^{4}(x^{2}-1)\right] ^{-1/2}dx,
\label{bbh1}
\end{equation}
which, with the substitution $u=1/x$, takes the form
\begin{equation}
I(x_{0})=2x_{0}\int_{0}^{1}
\left( u^{4}-x_{0}^{2}u^{2}+x_{0}^{2}-1\right) ^{-1/2}dx,
\label{bbh2}
\end{equation}
and it can be expressed as
\begin{equation}
I(x_{0})=\frac{2x_{0}}{\sqrt{x_{0}^{2}-1}}K\left( 
\frac{1}{\sqrt{x_{0}^{2}-1}}\right),
\label{bbh3}
\end{equation}
where $K(k)$ is the complete elliptic integral of the first kind\footnote{
$K(k)=\int _{0}^{\pi/2}(1-k^{2}\sin ^{2}\phi )^{-1/2}d\phi =\int _{0}^{1}
[(1-z^{2})(1-k^{2}z^{2})]^{-1/2}dz $} with argument $k=1/\sqrt{x_{0}^{2}-1}$ 
\cite{gradshteyn}. From Eq. (\ref{ps}), the photon sphere radius for the 
braneworld black hole is $x_{ps}=\sqrt{2}$. When $x_{0}$ takes 
values close to $x_{ps}$, $k$ takes values close to $1$, and for $0<1-k\ll 1$, 
$K(k)$ can be approximated by \cite{gradshteyn}
\begin{equation}
K(k)=\ln\left( \frac{4}{\sqrt{1-k^{2}}}\right), 
\label{bbh4}
\end{equation} 
then 
\begin{equation}
I(x_{0})=-\sqrt{2}\ln(x_{0}-\sqrt{2})+\sqrt{2}\ln 4,
\label{bbh5}
\end{equation} 
for $0<x_{0}-\sqrt{2}\ll 1$. The impact parameter $b$ (in units of $r_{H}$), 
defined as the perpendicular distance from the black hole to the asymptotic 
path at infinite, is more easily related with the lensing angles than the 
closest approach distance $x_{0}$. When the metric has the form of Eq. 
(\ref{met}), following Ref. \cite{weinberg}, the impact parameter is related 
to the closest approach distance by $b=[ h(x_{0})/f(x_{0})]^{1/2}$. Applying 
it to the braneworld black hole, the impact parameter is
\begin{equation}
b=\frac{x_{0}^{2}}{\sqrt{x_{0}^{2}-1}},
\label{b1b}
\end{equation}
which, making a second order Taylor expansion around $x_{0}=\sqrt{2}$, takes 
the form
\begin{equation}
b=2+2(x_{0}-\sqrt{2})^{2},
\label{bbh6}
\end{equation}
so, inverting this equation,
\begin{equation}
x_{0}-\sqrt{2}=\sqrt{\frac{b}{b_{ps}}-1},
\label{bbh7}
\end{equation}
where $b_{ps}=2$ is the critical impact parameter corresponding to 
$x_{0}=x_{ps}$. Replacing Eq. (\ref{bbh5}) in Eq. (\ref{alp1}) 
and using Eq. (\ref{bbh7}), the deflection angle is obtained 
as a function of the impact parameter $b$:
\begin{equation}
\alpha (b)=-c_{1}\ln \left( \frac{b}{b_{ps}}-1 \right) +c_{2}, 
\label{bbh8}
\end{equation}
where $c_{1}=\sqrt{2}/2$ and $c_{2}=\sqrt{2}\ln(4\sqrt{2})-\pi $. 
Eq. (\ref{bbh8}) represents the strong field limit approximation for the 
deflection angle produced by the braneworld black hole. Photons with an 
impact parameter slightly greater than the critical value $b_{ps}$ will spiral 
out, eventually reaching an observer after one or more turns around the black 
hole. In this case, the strong field limit gives a good approximation for the 
deflection angle (see discussion in Refs. \cite{bozza1,bozman2}). Those 
photons whose impact parameter is smaller than $b_{ps}$ will spiral into the 
black hole, not reaching any observer outside the photon sphere.

\section{Positions and magnifications of the relativistic images}\label{s3}

In this section the positions and magnifications of the relativistic images in 
the strong field limit are calculated for the braneworld black hole, first for 
a point source and then for a spherical extended source.

\subsection{Point source}

\begin{figure}[t!]
\vspace{0cm} 
\includegraphics[width=6.4cm]{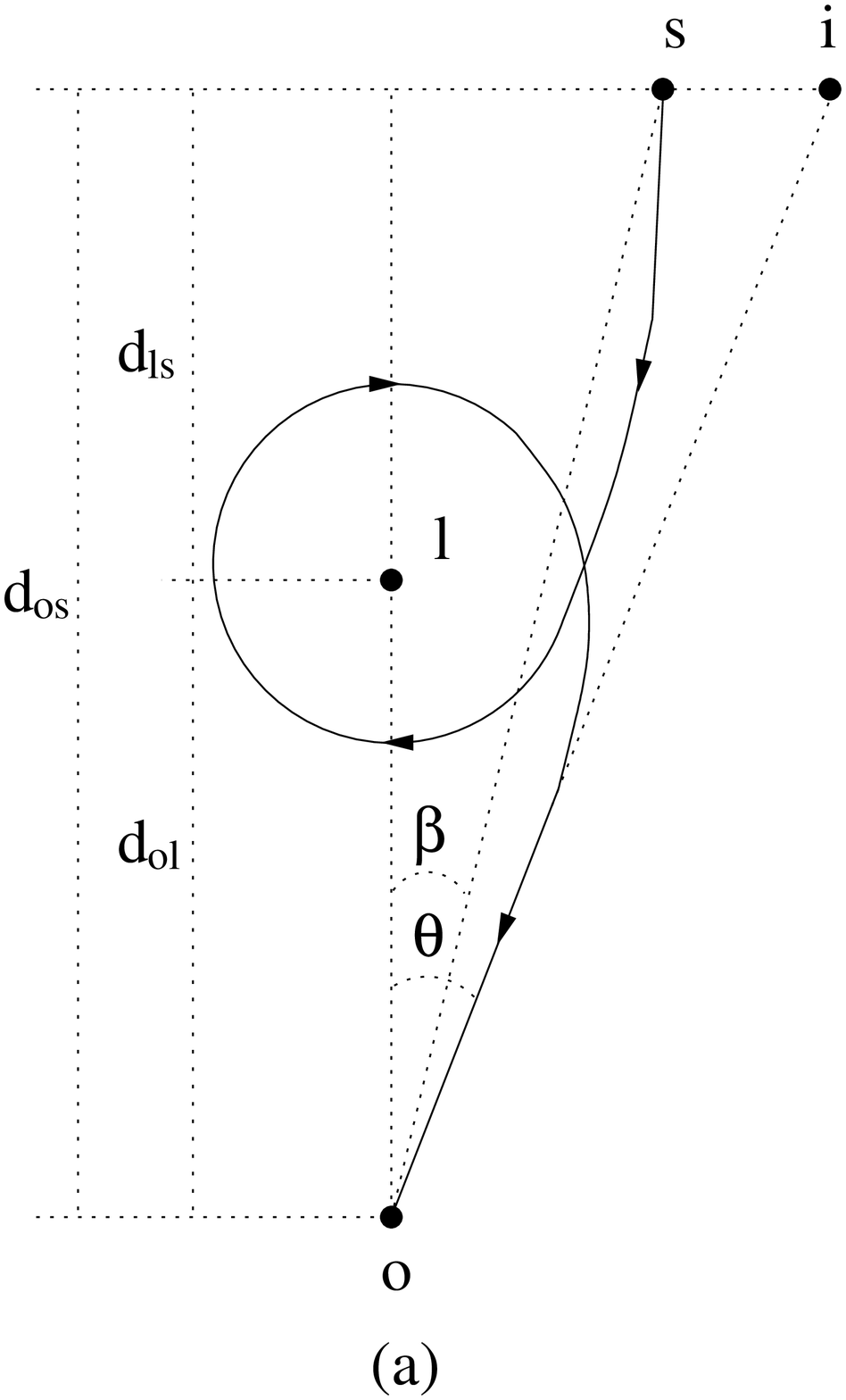} 
\hspace{-1.3cm}
\includegraphics[width=6.4cm]{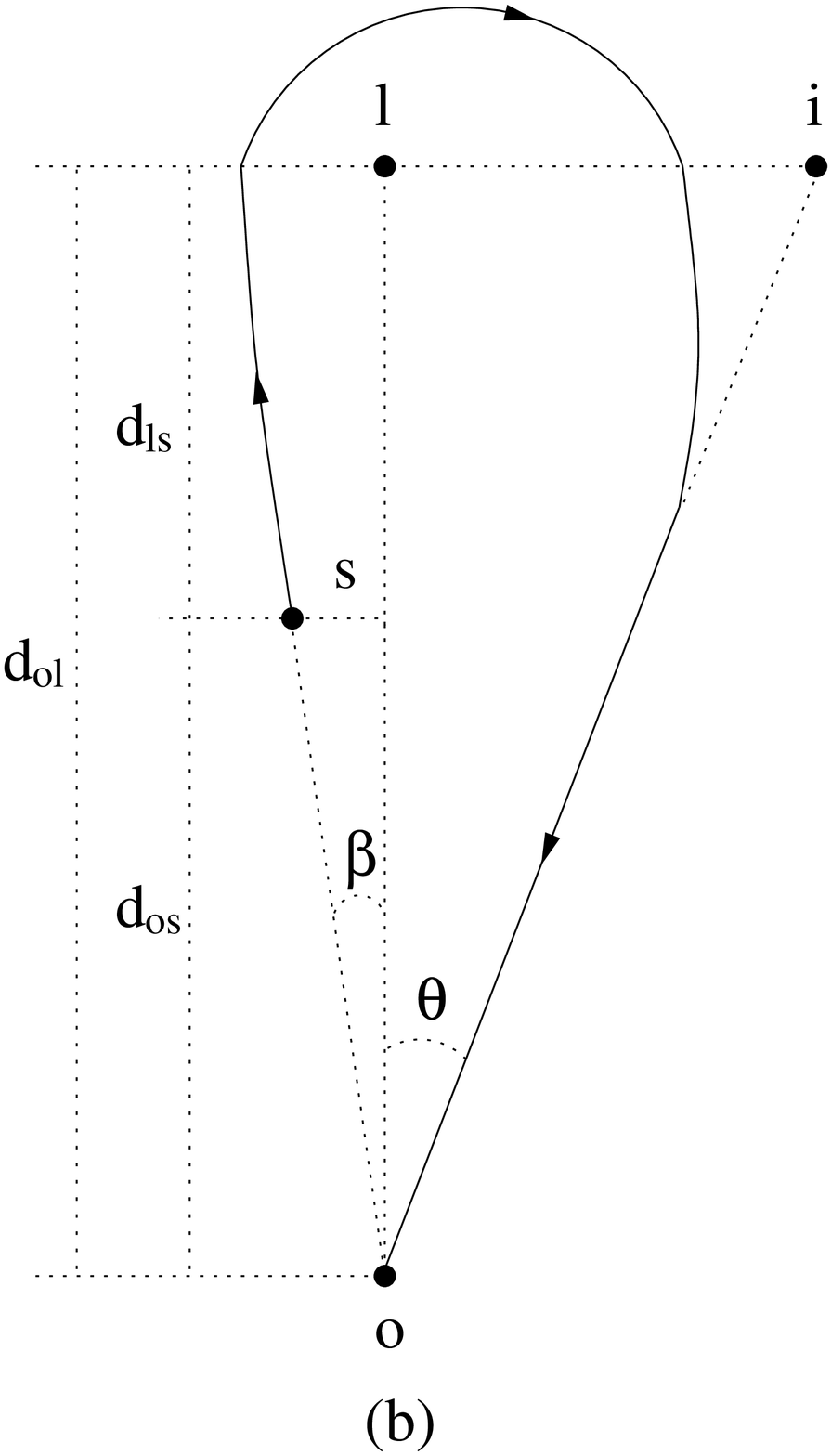}
\hspace{-1.3cm}
\includegraphics[width=6.4cm]{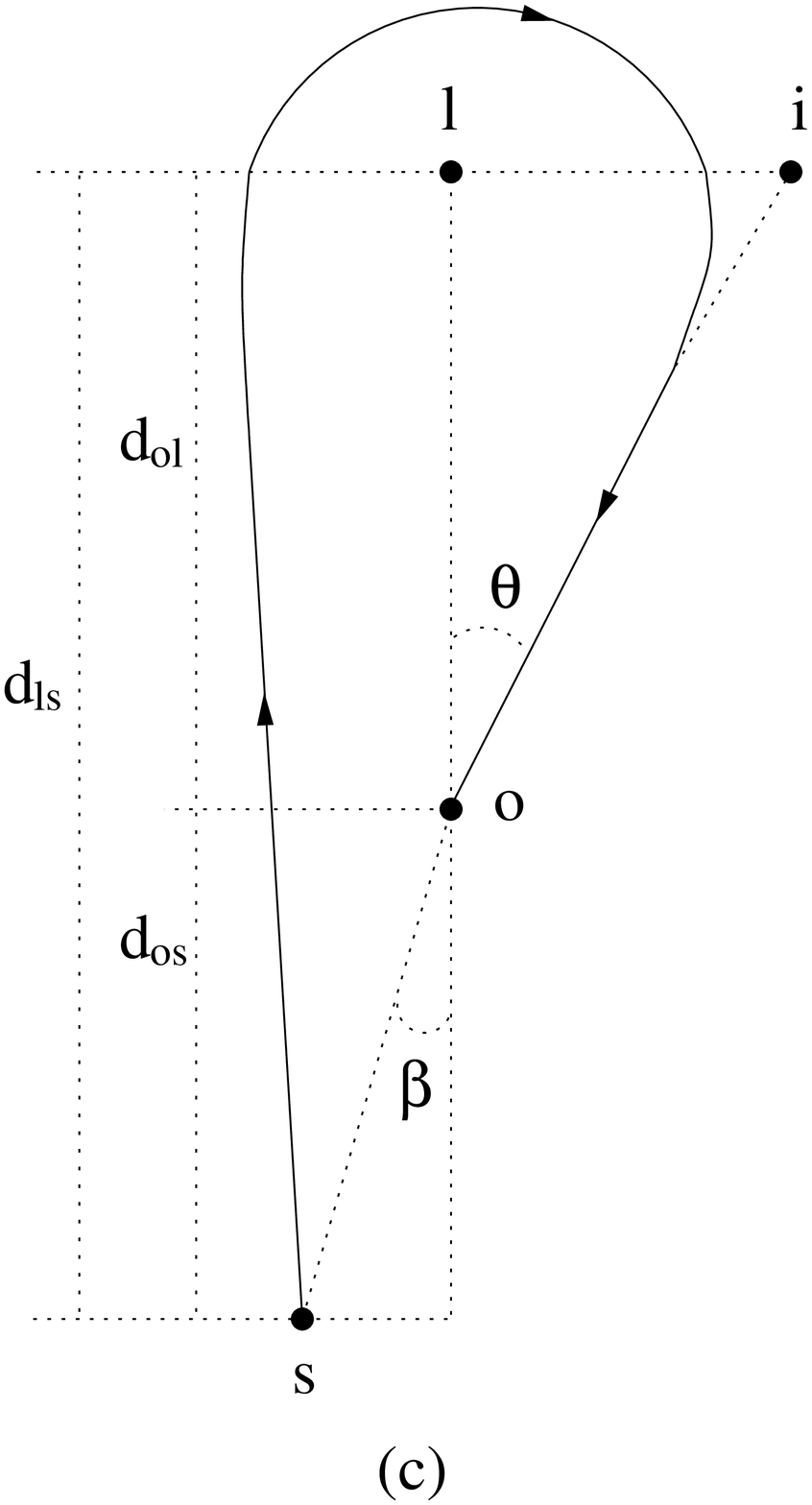}
\vspace{-0.5cm}
\caption{Schematic representation of the lens geometries. Case (a) corresponds 
to standard lensing, and cases (b) and (c) to retrolensing. The source (s), 
the lens (l), the observer (o) and the first relativistic image (i) are shown.}
\label{f1}
\end{figure}

The lensing scenario, shown in Fig. \ref{f1}, consists of a point source of 
light (s), an observer (o), and the braneworld black hole, which it is called 
the lens (l). The line joining the observer and the lens define the optical 
axis. The background space-time is considered asymptotically flat, with the 
observer and the source immersed in the flat region. The angular position of 
the source is $\beta $ and the angular position of the images (i) is 
$\theta $, both seen from the observer. The observer-source, observer-lens 
and the lens-source distances, here taken much greater than the horizon 
radius, are (in units of the horizon radius), $d_{os}=D_{os}/r_{H}$, 
$d_{ol}=D_{ol}/r_{H}$ and $d_{ls}=D_{ls}/r_{H}$, respectively. There are three 
possible configurations: (a) the lens between the observer and the source, 
(b) the source between the observer and the lens, and (c) the observer between 
the source and the lens. The situation (a) is called standard lensing and 
both (b) and (c) are called retrolensing. It can be taken $\beta >0$ without 
losing generality. The lens equation is:
\begin{equation}
\tan \beta =\tan \theta -c_{3}\left[ \tan (\alpha -\theta)
+\tan \theta \right] ,
\label{le1}
\end{equation}
where $c_{3}=d_{ls}/d_{os}$ \cite{virbha1} for standard lensing (a), and 
$c_{3}=d_{os}/d_{ol}$ \cite{eitor} and $c_{3}=d_{os}/d_{ls}$ for the cases 
(b) and (c) of retrolensing, respectively. The lensing effects are more 
prominent when the objects are highly aligned (for a discussion see, for 
instance, Ref. \cite{bozman2}). In this case, the angles $\beta $ 
and $\theta $ are small and $\alpha $ is closer to a multiple of $\pi $. Two 
infinite sets of relativistic images are formed. For the first set of images  
(see Fig. \ref{f1}), the deflection angle can be 
written as $\alpha =m\pi +\Delta \alpha _{m}$, with $m\in \mathbb{N}$ ($m$ is 
even for standard lensing and odd for retrolensing) and 
$0<\Delta \alpha _{m}\ll 1$. Then, the lens equation takes the form
\begin{equation}
\beta =\theta -c_{3}\Delta \alpha _{m}.
\label{le2}
\end{equation}
To obtain the other set of images, it should be taken 
$\alpha =-m\pi -\Delta \alpha _{m}$, so $\Delta \alpha _{m}$ must
be replaced by $-\Delta \alpha _{m}$ in Eq. (\ref{le2}). In the case of
perfect alignment, an infinite sequence of concentric Einstein rings is
obtained.\\

From the lens geometry it is easy to see that
\begin{equation}
b=d_{ol}\sin \theta ,
\label{b2}
\end{equation}
which can be approximated to first order in $\theta $ by 
$b=d_{ol}\theta $, so the deflection angle given by Eq. (\ref{bbh8}) can 
be written as a function of $\theta $:
\begin{equation}
\alpha (\theta )=-c_{1}\ln \left( \frac{\theta}{\theta _{ps}}-1 \right) 
+c_{2}, 
\label{bbh9}
\end{equation}
with $\theta _{ps}=2/d_{ol}$.
Inverting Eq. (\ref{bbh9}) to obtain $\theta (\alpha )$
\begin{equation}
\theta (\alpha )=\theta _{ps}\left[ 1+e^{(c_{2}-\alpha )/c_{1}}\right] ,
\label{bbh10}
\end{equation}
and making a first order Taylor expansion around $\alpha =m\pi $, the angular 
position of the $m$-th image can be approximated by
\begin{equation}
\theta _{m}=\theta ^{0}_{m}-\zeta _{m}\Delta \alpha _{m},
\label{bbh11}
\end{equation}
with
\begin{equation}
\theta ^{0}_{m}=\theta _{ps}\left[ 1+e^{(c_{2}-m\pi )/c_{1}}
 \right] ,
\label{bbh12}
\end{equation}
and
\begin{equation}
\zeta _{m}=\frac{\theta _{ps}}{c_{1}}e^{(c_{2}-m\pi )/c_{1}}.
\label{bbh13}
\end{equation}
From Eq. (\ref{le2})
\begin{equation}
\Delta \alpha _{m}=\frac{\theta _{m}-\beta }{c_{3}},
\label{bbh14}
\end{equation}
and replacing it in Eq. (\ref{bbh11}) leads to
\begin{equation}
\theta _{m}=\theta ^{0}_{m}-\frac{\zeta _{m}}{c_{3}}(\theta _{m}-\beta ),
\label{bbh15}
\end{equation}
which can be written in the form
\begin{equation}
\left( 1+\frac{\zeta _{m}}{c_{3}}\right)\theta _{m}=\theta ^{0}_{m}+
\frac{\zeta _{m}}{c_{3}}\beta ,
\label{bbh16}
\end{equation}
then
\begin{equation}
\theta _{m}=\left( 1+\frac{\zeta _{m}}{c_{3}}\right) ^{-1}\left( 
\theta ^{0}_{m}+\frac{\zeta _{m}}{c_{3}}\beta \right) ,
\label{bbh16a}
\end{equation}
which, using that $0<\zeta _{m}/c_{3}\ll 1$, can be approximated by
\begin{equation}
\theta _{m}=\left( 1-\frac{\zeta _{m}}{c_{3}}\right)\left( \theta ^{0}_{m}+
\frac{\zeta _{m}}{c_{3}}\beta \right) .
\label{bbh16b}
\end{equation}
Then, keeping only the first order term in $\zeta _{m}/c_{3}$, the angular 
positions of the images are finally given by
\begin{equation}
\theta _{m}=\theta ^{0}_{m}+\frac {\zeta _{m}}{c_{3}}(\beta -\theta ^{0}_{m}).
\label{bbh17}
\end{equation}
The second term in Eq. (\ref{bbh17}) is a small correction on 
$\theta ^{0}_{m}$, so all images lie very close to $\theta ^{0}_{m}$. With a 
similar treatment, the other set of relativistic images have angular 
positions 
\begin{equation}
\theta _{m}=-\theta ^{0}_{m}+\frac {\zeta _{n}}{c_{3}}(\beta +\theta ^{0}_{m}).
\label{bbh18}
\end{equation}
When $\beta =0$ an infinite sequence of Einstein rings is formed,  with 
angular radius
\begin{equation}
\theta ^{E}_{m}=\left( 1-\frac {\zeta _{m}}{c_{3}}\right) \theta ^{0}_{m}.
\label{bbh19}
\end{equation}
\\
Since gravitational lensing conserves surface brightness \cite{schne}, the 
ratio of the solid angles subtended by the image and the source gives the 
amplification of the $m$-th image:
\begin{equation}
\mu _{m}=\left| \frac{\sin \beta }{\sin \theta _{m}}
\frac{d\beta }{d\theta _{m}}\right|^{-1}\approx 
\left| \frac{\beta }{\theta _{m}} \frac{d\beta }{d\theta _{m}}\right|^{-1},
\label{mu1}
\end{equation}
so, using Eq. (\ref{bbh17}),
\begin{equation}
\mu _{m}=\frac{1}{\beta}\left[ \theta ^{0}_{m}+
\frac {\zeta _{m}}{c_{3}}(\beta - \theta ^{0}_{m})\right]
\frac {\zeta _{m}}{c_{3}},
\label{mu2}
\end{equation}
which can be approximated to first order in $\zeta _{m}/c_{3}$ by
\begin{equation}
\mu _{m}=\frac{1}{\beta}\frac{\theta ^{0}_{m}\zeta _{m}}{c_{3}}.
\label{mu3}
\end{equation}
The same result is obtained for the other set of images. The first
relativistic image is the brightest one, and the magnifications
decrease exponentially with $m$. \\

For standard lensing $m=2n$, and the total magnification, considering both 
sets of images, is $\mu =2\sum\limits_{n=1}^{\infty }\mu _{n}$, which using  
Eqs. (\ref{bbh12}), (\ref{bbh13}) and (\ref{mu3}), leads to
\begin{equation}
\mu =\frac {1}{\beta}\frac {8}{d_{ol}^{2}c_{1}c_{3}}
\frac{e^{c_{2}/c_{1}}\left( 1+e^{c_{2}/c_{1}}+e^{2\pi /c_{1}}\right) }
{e^{4\pi /c_{1}}-1}.
\label{mu4}
\end{equation}
For retrolensing, $m=2n-1$, the total magnification, considering both 
sets of images, is 
\begin{equation}
\mu =\frac {1}{\beta}\frac {8}{d_{ol}^{2}c_{1}c_{3}}
\frac{e^{(c_{2}+\pi )/c_{1}}\left[ 1+e^{(c_{2}+\pi )/c_{1}}+e^{2\pi /c_{1}}
\right] }{e^{4\pi /c_{1}}-1}.
\label{mu5}
\end{equation}
Note that the amplifications of the strong field images are greater for 
retrolensing. In both cases, the magnifications are proportional to 
$1/d_{ol}^{2}$, which is a very small factor. Then, the relativistic images 
are very faint, unless $\beta $ has values close to zero, i.e. nearly perfect 
alignment. For $\beta =0$, the amplification becomes infinite, and the point 
source approximation breaks down, so an extended source analysis is necessary.

\subsection{Extended source}

For an extended source, it is necesary to integrate over its luminosity 
profile to obtain the magnification of the images:
\begin{equation}
\mu =\frac{\int\!\!\int_{S}\mathcal{I}\mu _{p}dS}{\int\!\!\int_{S}
\mathcal{I}dS}, \label{ext1}
\end{equation}
where $\mathcal{I}$ is the surface intensity distribution of the source and
$\mu _{p}$ is the magnification corresponding to each point of the source.
If the source is an uniform disk $D(\beta _{c},\beta _{s})$, with angular 
radius $\beta _{s}$ and centered in $\beta _{c}$ (taken positive), 
Eq. (\ref{ext1}) can be put in the form
\begin{equation}
\mu =\frac{\int\!\!\int_{D(\beta _{c},\beta _{s})}\mu _{p} dS}
{\pi \beta _{s}^{2}}.
\label{ext2}
\end{equation}
Then, using Eq. (\ref{mu3}), the magnification of the relativistic $m$-th 
image (with $m$ even for standard lensing and odd for retrolensing) for an 
extended uniform source is
\begin{equation}
\mu _{m}=\frac{I}{\pi \beta _{s}^{2}}\frac {\theta ^{0}_{m}\zeta _{m}}{c_{3}},
\label{ext3}
\end{equation}
with $I=\int\!\!\int_{D(\beta _{c},\beta _{s})} (1/\beta )dS$. This
integral can be calculated in terms of elliptic integrals:
\begin{equation}
I=2\left[ (\beta _{s}+\beta _{c})
E\left(\frac{2\sqrt{\beta _{s}\beta _{c}}}{\beta _{s}+\beta _{c}}\right)
+(\beta _{s}-\beta _{c})K\left(\frac{2\sqrt{\beta _{s}\beta _{c}}}
{\beta _{s}+\beta _{c}}\right) \right] ,
\label{ext6}
\end{equation}
where $K(k)$ and $E(k)$ are respectively, complete elliptic integrals of the 
first and second kind\footnote{
$E(k)=\int_{0}^{\pi /2}\left( 1-k^2\sin ^{2}\phi \right) ^{1/2}d\phi =
\int _{0}^{1}(1-z^{2})^{-1/2}(1-k^{2}z^{2})^{1/2}dz $} with 
argument $k=2\sqrt{\beta _{s}\beta _{c}}/(\beta _{s}+\beta _{c})$ 
\cite{gradshteyn}. Then the total amplification for standard lensing is
\begin{equation}
\mu =\frac{I}{\pi \beta _{s}^{2}}\frac {8}{d_{ol}^{2}c_{1}c_{3}}
\frac{e^{c_{2}/c_{1}}\left( 1+e^{c_{2}/c_{1}}+e^{2\pi /c_{1}}\right) }
{e^{4\pi /c_{1}}-1},
\label{ext4}
\end{equation}
and for retrolensing is
\begin{equation}
\mu =\frac{I}{\pi \beta _{s}^{2}}\frac {8}{d_{ol}^{2}c_{1}c_{3}}
\frac{e^{(c_{2}+\pi )/c_{1}}\left[ 1+e^{(c_{2}+\pi )/c_{1}}+e^{2\pi /c_{1}}
\right] }{e^{4\pi /c_{1}}-1}.
\label{ext5}
\end{equation}
These expresions always give finite magnifications, even in the case of 
complete alignment.

\section{Comparison with Schwarzschild black holes}\label{s4}

The equations that give the positions and magnifications of the relativistic 
images obtained in Sec. \ref{s3} can be applied to the four dimensional 
Schwarzschild black hole, with the distances measured in units of the 
Schwarzschild radius $r_{H}^{Schw}=2(M/M_{4})l_{4}$, the photon sphere radius 
given by $r_{ps}^{Schw}=3r_{H}^{Schw}/2$, and the constants $c_{1}$ and 
$c_{2}$ replaced by $c_{1}^{Schw}=1$ and 
$c_{2}^{Schw}=\ln [216(7-4\sqrt{3})]-\pi $ \cite{bozza2} ($c_{3}$ does not 
change). The quotient between the $m$-th Einstein radius of the braneworld 
black hole and that of the Schwarzschild black hole with the same mass can be
expressed in the form
\begin{equation}
\frac{\theta^{E}_{m}}{\theta^{E,Schw}_{m}}=\chi (m)
\frac{c_{3}D_{ol}-32^{\sqrt{2}}2\sqrt{2}e^{-(m+1)\sqrt{2}\pi }r_{H}}
{c_{3}D_{ol}-324(7-4\sqrt{3})e^{-(m+1)\pi }r_{H}^{Schw}}
\left( \frac{l}{l_{4}}\right)^{1/2}\left( \frac{M}{M_{4}}\right)^{-1/2},
\label{comp1a}
\end{equation}
with $m$ even for standard lensing and odd for retrolensing, and
\begin{equation} 
\chi (m)=\sqrt{\frac{32}{27\pi }}\frac{1+32^{\sqrt{2}}e^{-(m+1)\sqrt{2}\pi }}
{1+216(7-4\sqrt{3})e^{-(m+1)\pi }}.
\label{comp1b}
\end{equation}
In the usual case when $c_{3}D_{ol}\gg 32^{\sqrt{2}}2\sqrt{2}r_{H}$ and  
$c_{3}D_{ol}\gg 324(7-4\sqrt{3})r_{H}^{Schw}$, Eq. (\ref{comp1a}) can be 
approximated by
\begin{equation}
\frac{\theta^{E}_{m}}{\theta^{E,Schw}_{m}}=\chi (m)
\left( \frac{l}{l_{4}}\right)^{1/2}\left( \frac{M}{M_{4}}\right)^{-1/2}.
\label{comp1c}
\end{equation}
For the first Einstein radius, Eq. (\ref{comp1b}) gives $\chi \approx 0.6136$ 
for standard lensing, and $\chi \approx 0.6080$ for retrolensing. The 
quotient of the magnifications for the $m$-th image can be written as
\begin{equation}
\frac{\mu _{m}}{\mu ^{Schw}_{m}}=\omega (m)\frac{l}{l_{4}}\frac{M_{4}}{M},
\label{comp2a}
\end{equation}
with
\begin{equation}
\omega (m)=\frac{32^{\sqrt{2}}4\sqrt{2}}{729(7-4\sqrt{3})\pi } 
\frac{\left[ 1+32^{\sqrt{2}}e^{-(m+1)\sqrt{2}\pi }\right] 
e^{-(m+1)(\sqrt{2}-1)\pi }}{1+216(7-4\sqrt{3})e^{-(m+1)\pi }},
\label{comp2b}
\end{equation}
which gives for the first image the values $\omega \approx 0.09318$ for 
standard lensing, and $\omega \approx 0.3392$ for retrolensing. The ratio of 
the total magnifications is
\begin{equation}
\frac{\mu }{\mu ^{Schw}}=\kappa \frac{l}{l_{4}}\frac{M_{4}}{M},
\label{comp3a}
\end{equation}
with
\begin{equation}
\kappa =\frac{32^{\sqrt{2}}\sqrt{2}\left( e^{4\pi }-1\right) 
e^{-(\sqrt{2}-1)\pi }\left( 1+32^{\sqrt{2}}e^{-\sqrt{2}\pi }+
e^{2\sqrt{2}\pi }\right) }
{324(7-4\sqrt{3})\left( e^{4\sqrt{2}\pi }-1\right) 
\left[ 1+216(7-4\sqrt{3})e^{-\pi }+e^{2\pi }\right] \pi } \approx 0.05232,
\label{comp3b}
\end{equation}
for standard lensing, and
\begin{equation}
\kappa =\frac{32^{\sqrt{2}}\sqrt{2}\left( e^{4\pi }-1\right) 
\left( 1+32^{\sqrt{2}}+e^{2\sqrt{2}\pi }\right) }
{324(7-4\sqrt{3})\left( e^{4\sqrt{2}\pi }-1\right) 
\left[ 1+216(7-4\sqrt{3})+e^{2\pi }\right] \pi }\approx 0.1905,
\label{comp3c}
\end{equation}
for retrolensing.\\

The quotient of the Einstein radii is roughly proportional to 
$r_{H}/r^{Schw}_{H}\propto (l/M)^{1/2}$, whereas the ratio of the 
amplifications is proportional to $(r_{H}/r^{Schw}_{H})^{2}\propto l/M$. Note 
that even in the case of a braneworld black hole and a Schwarzschild 
black hole with the same mass and the same horizon event radius, the Einstein 
radii and magnifications are different. As in the weak field lensing case 
\cite{majumuk}, the strong field images for the braneworld black hole increase 
their brightness with respect to the Schwarzschild black hole for larger 
values of the extra dimension $l$. 

\section{Conclusions}\label{s5}

In this work, the positions and magnifications of the relativistic images were 
calculated, using the strong field limit, for a braneworld black hole. These 
black holes were previously studied as primordial black holes 
\cite{cgl,majumdar}, and can be relics of a past high energy phase of the 
Universe. They have a larger size, are colder, and live longer than their 
four dimensional Schwarzschild counterpart of the same mass \cite{kanti}. 
Current experiments that test the gravitational inverse square law, lead to an 
upper limit for the extra dimension given by $l\le 0.1$ mm \cite{long}. The 
condition that the event horizon radius $r_{H}$ is much smaller than the Ads 
radius $l$ for the braneworld black holes studied in this paper, means that 
these black holes, if they exist, are very small. The astronomical observation 
of the relativistic images is a difficult task that is beyond current 
technologies, and it will be a challenge for the next generation of 
instruments in the case of more massive black holes \cite{bozman2}. The 
observation seems even more difficult in the case of the small 
size braneworld black holes analyzed in this paper. But there 
could be another route for seeing the strong field 
lensing effects for these low mass black holes. The presence of the extra 
dimension dramatically decreases the energy necessary to produce black holes 
by particle collisions \cite{kanti,lab}. Only energies of about $1$ TeV 
are needed instead of energy scales about $10^{16}$ TeV required if no extra 
dimensions are present. These small size black holes could be created in the 
next generation particle accelerators or detected in cosmic rays 
\cite{kanti,lab}. With these black holes acting as gravitational lenses, 
the order of magnitude of the distances involved would be meters or less, 
instead of kilo-parsecs, so that the angular positions and magnifications 
obtained in Sec. \ref{s3} will be considerably larger than in the astronomical 
case corresponding to black holes with the same mass \footnote{For example, 
let us consider, in the standard lensing configuration, a black hole lens 
with $r_{H}=10^{-3}$ mm placed halfway between a point source and 
the observer in two cases: a laboratory (L) situation with $D_{os}^{(L)}=1$ m 
and an astronomical (A) situation with $D_{os}^{(A)}=10$ kpc. Then,
the quotients between the first Einstein angular radii and the total 
magnifications are, respectively, 
$\theta^{E\,(L)}_{1}/\theta^{E\,(A)}_{1}=3.1\times 10^{20}$ and 
$\mu^{(L)}/\mu^{(A)}=9.5\times 10^{40}$.}. If the extra dimensions hypothesis 
is correct, it will open up the exciting possibility that the interesting 
phenomena of strong field lensing can be observed in the laboratory 
in future times. 

\section*{Acknowledgements}

This work has been partially supported by UBA (UBACYT X-103).


\end{document}